\documentclass[fleqn,10pt]{wlscirep}
\usepackage[utf8]{inputenc}
\usepackage[T1]{fontenc}

\usepackage{ragged2e}
\usepackage{graphicx}
\usepackage{float}
\usepackage{amsthm}
\usepackage{amsmath} 
\usepackage{amsfonts}
\usepackage{dsfont} 
\usepackage{amssymb}
\usepackage{algorithm}
\usepackage{relsize}
\usepackage{algorithm,algpseudocode}
\usepackage{breqn}
\usepackage{etoolbox}
\usepackage{lmodern}
\usepackage{tikz}
\usepackage[font=footnotesize]{caption}
\usepackage{subcaption}
\usepackage{letltxmacro}
\usepackage{scalerel}
\usepackage[toc,page]{appendix}
\usepackage{tabularx,booktabs}
\usepackage{blindtext}
\usepackage{enumitem}
\usepackage{titlesec}
\usepackage{multirow}
\usepackage{stackrel}
\usepackage{adjustbox}
\usepackage{mathtools} 
\mathtoolsset{showonlyrefs}

\graphicspath{ {images/} }
\DeclareRobustCommand{\c}[1]{\mathcal{#1}}

\DeclareRobustCommand{\t}[1]{\text{#1}}

\DeclareRobustCommand{\Rb}{\mathbb{R}}

\DeclareRobustCommand{\bZ}{\boldsymbol{Z}}

\DeclareRobustCommand{\bX}{\boldsymbol{X}}

\DeclareRobustCommand{\bOmega}{\boldsymbol{\Omega}}
\newcommand{\hnu}{\widehat{\nu}} 

\newcommand{\bI}{\boldsymbol{I}}

\newcommand{\bY}{\boldsymbol{Y}}

\DeclareRobustCommand{\bhU}{\widehat{\boldsymbol{U}}}
\DeclareRobustCommand{\bhV}{\widehat{\boldsymbol{V}}}

\DeclareRobustCommand{\bhSigma}{\widehat{\boldsymbol{\Sigma}}}
\DeclareRobustCommand{\Ac}{\mathcal{A}}

\DeclareRobustCommand{\Kc}{\mathcal{K}}

\DeclareRobustCommand{\Pc}{\mathcal{P}}

\newcommand{\LFc}{\mathcal{LF}}

\newcommand{\Tc}{\mathcal{T}}

\DeclareRobustCommand{\Sc}{\mathcal{S}}
\DeclareRobustCommand{\Ac}{\mathcal{A}}

\DeclareRobustCommand{\Rb}{\mathbb R}

\DeclareMathOperator*{\argmin}{\arg\!\min}

\newcommand{\distas}[1]{\mathbin{\overset{#1}{\kern\z@\sim}}}%
\newsavebox{\mybox}\newsavebox{\mysim}
\newcommand{\distras}[1]{%
  \savebox{\mybox}{\hbox{\kern3pt$\scriptstyle#1$\kern3pt}}%
  \savebox{\mysim}{\hbox{$\sim$}}%
  \mathbin{\overset{#1}{\kern\z@\resizebox{\wd\mybox}{\ht\mysim}{$\sim$}}}%
}

\newcommand{\hEx}{\widehat{\Ex}}
\newcommand{\hbeta}{\widehat{\beta}}

\newcommand{\Ex}{\mathbb{E}}

\newcommand\independent{\protect\mathpalette{\protect\independenT}{\perp}}
\def\independenT#1#2{\mathrel{\rlap{$#1#2$}\mkern2mu{#1#2}}}

\newcommand{\hu}{\widehat{u}}
\newcommand{\hv}{\widehat{v}} 
\newcommand{\hsigma}{\widehat{\sigma}} 

\newcommand{\Matching}{\texttt{Matching}}
\newcommand{\SNN}{\texttt{SNN}}
\newcommand{\LOCF}{\texttt{LOCF}}
\newcommand{\Naive}{\texttt{Na\"{i}ve}}

\newcommand{\PCR}{\texttt{PCR}}

\newcommand{\obs}{\text{obs}}

\title{Personalized Predictions from Population Level Experiments: 
A Study on Alzheimer's Disease}

\author[1,*]{Dennis Shen}
\author[2]{Anish Agarwal}
\author[3]{Vishal Misra}
\author[4]{Bjoern Schelter}
\author[5]{Devavrat Shah}
\author[4]{Helen Shiells}
\author[4]{Claude Wischik} 
\affil[1]{USC, Department of Data Sciences \& Operations}
\affil[2]{Columbia University, Department of Industrial Engineering \& Operations Research}
\affil[3]{Columbia University, Department of Computer Science} 
\affil[4]{TauRx Therapeutics}
\affil[5]{MIT, Department of Electrical Engineering \& Computer Science} 
\affil[*]{dennis.shen@marshall.usc.edu}


\begin{abstract}
The purpose of this article is to infer patient level outcomes from population level randomized control trials (RCTs). 
In this pursuit, we utilize the recently proposed synthetic nearest neighbors (\SNN) estimator. 
At its core, \SNN~leverages information across patients to impute missing data associated with each patient of interest. 
We focus on two types of missing data: 
(i) unrecorded outcomes from discontinuing the assigned treatments 
and
(ii) unobserved outcomes associated with unassigned treatments.
Data imputation in the former powers and de-biases RCTs, 
while data imputation in the latter simulates ``synthetic RCTs'' to predict the outcomes for each patient under every treatment. 
The \SNN~estimator is interpretable, transparent, and causally justified under a broad class of missing data scenarios. 
Relative to several standard methods, we empirically find that \SNN~performs well for the above two applications using Phase 3 clinical trial data on patients with Alzheimer's Disease. 
Our findings directly suggest that \SNN~can tackle a current pain point within the clinical trial workflow on patient dropouts and serve as a new tool towards the development of precision medicine. 
Building on our insights, we discuss how \SNN~can further generalize to real-world applications. 
\end{abstract}

\begin{document}

\flushbottom
\maketitle
\thispagestyle{empty}

% INTRODUCTION
\section*{Introduction} \label{sec:intro} 
Randomized controlled trials (RCTs) remain the gold standard mechanism to draw causal conclusions about the safety and efficacy of new treatments. 
Typically, inferences of RCTs are drawn at the population level, e.g., estimating average treatment effects (ATEs) across all participating patients. 
However, patients, particularly those with complex disorders, are often heterogeneous. 
That is, two patients with the same disease may exhibit different patterns of disease progression and respond differently to the same treatment. 
Herein lies the fundamental problem of ATEs: they are insufficient to describe the distribution of treatment effects across patients.
This realization has sparked the rise of precision medicine (also known as personalized medicine), which aims to understand disease treatment and prevention strategies that factor in patient variability in genes, behaviors, and environment \cite{precisionmed}. 
The penultimate promise of this initiative is to match the right treatments to the right patients. 

To this end, there has been a wave of innovation in estimating heterogeneous treatment effects. 
Methodologically, there are two key approaches: (i) conditional average treatment effect (CATE) function estimation \cite{cate0, cate1, cate2, cate3, stadisc}, and (ii) subgroup analysis \cite{sg1, sg2, sg3}.
Broadly speaking, both approaches focus on estimating subgroup level effects from standard trial data. 
The challenge for current methods in both categories is that budget and resource constraints often restrict RCTs to be only sufficiently powered to estimate population level quantities such as the ATE. 
Therefore, there may be insufficient data and signal strength within each patient subgroup for existing techniques to detect statistically significant effects. 
On the experimental front, the advancement of biomarker identification has catapulted the adoption of biomarker-guided trials such as basket trials and umbrella trials that group patients by their characteristics \cite{basket}. 
Biomarker-guided trials are more targeted compared to traditional trials, but are often highly complex and present unique challenges to patients, investigators, regulatory agencies, and industry \cite{Cecchini2049}. 
%

%Despite the tremendous progress made in the past decade or so in estimating heterogenous treatment effects, 
To date, estimating individual treatment effects has remained largely elusive. 
%modern approaches fall short of estimating individual treatment effects (ITEs).  
%
In order to make predictions at the patient level granularity, researchers would ideally test each treatment on every patient to observe all therapeutic outcomes. 
%researchers would ideally have access to each patient's outcomes under every treatment scenario.
%
Experimentally, this is simply infeasible. 
Patients can only be assigned a single treatment and even then, their outcomes under the assigned treatment are commonly only partially observed given the high dropout rate in clinical trials \cite{dropout1}.
The unrecorded outcomes induced by dropouts compromise inferences from RCTs since they decrease power and inflate biases \cite{missingdata1}.

\subsection*{Imputing missing data via tensor completion}
This article looks to provide both a statistical framework and methodology to predict patient level outcomes from (sub)population level trials. 
Towards this objective, we build upon the framework introduced in \cite{si} and encode our observations into an order-three tensor indexed by patients, time, and treatments. 
The $(i,t,a)$th entry of the tensor corresponds the potential clinical outcome of patient $i$ at visit $t$ under assignment $a$. 
From this perspective, a fully observed tensor reveals the personalized health trajectory for each patient under every treatment, thus achieving precision medicine. 
Reality, however, is not so picturesque. 
Researchers can only observe a sparse subset of the tensor: (i) if a patient withdraws from the trial, then their subsequent outcomes are unrecorded; (ii) if a patient is assigned to one treatment arm then their outcomes under other treatments are unobserved. 
Through this lens, we recast the problem of imputing missing data as one of tensor completion. 
A challenge in this undertaking is that the entries can be  ``missing not at random'' (MNAR), i.e., the underlying values of the tensor and the missingness mechanism are  driven by common {\em observed} and {\em unobserved} factors. %, which we call latent confounders. 
%

% SNN 
\subsection*{\SNN: tensor completion via synthetic matching}
We utilize the recently proposed synthetic nearest neighbors (\SNN) estimator of \cite{snn} to complete the tensor. 
At a high level, the \SNN~estimator imputes missing values in the tensor by leveraging information across patients whose outcomes are observed.  

To illustrate \SNN, suppose Alice withdraws from the trial before its completion so her remaining outcomes in the trial are unrecorded. 
One popular class of imputation methods are those based on matching.  
In principle, matching methods first identify the $K \ge 1$ patients that stayed throughout the trial who are most similar to Alice and average their outcomes as a proxy for Alice's outcomes had she not dropped out. 
Given the heterogeneity amongst patients, it is unlikely for any patient within the trial to emulate Alice precisely. 
This motivates the \SNN~approach.
\SNN~is similar in spirit to matching, but rather than identifying a collection of patients that {\em each} behave like Alice, \SNN~identifies a collection of patients that {\em collectively} behaves like Alice. 
Specifically, \SNN~creates a ``synthetic'' Alice from a weighted composition of actual patients whose pretreatment variables and outcome variables most closely mirrors that of Alice's prior to her withdrawal. 
The contribution of each patient in the construction of a synthetic Alice can be easily seen from his or her associated weight. 
If a patient perfectly resembles Alice, then all of the mass would be placed on this patient and \SNN~reduces to standard matching.  
In this view, \SNN~enjoys the same advantages as matching in offering interpretability and transparency, but allows for increased prediction capability given its increased model flexibility.

%As another illustration, suppose 
Now, suppose our interest is to provide treatment recommendations for Bob. 
Traditional one-size-fits all approaches recommend the treatment with the highest ATE; 
hence, Bob would receive the same recommendation as every other patient. 
More refined metrics such as the CATE can also be applied, though these recommendations are still determined at the subgroup level; here, Bob would receive the same recommendation as every patient who shares similar characteristics with him.
In contrast, \SNN~offers a personalized recommendation for Bob by providing a prediction for Bob's unique outcomes under every  treatment. 
Similar to the previous example with Alice, for every unassigned treatment, \SNN~constructs a neighborhood of synthetic Bob-like patients via a weighted combination of existing patients within the treatment arm. 
The associated outcomes are then further averaged to predict Bob's counterfactual outcomes under said treatment. 
More generally, \SNN~can utilize standard (sub)population level RCT data, where each patient is assigned a single treatment, to predict the outcome for the remaining untested patient-treatment combinations. 
In doing so, it augments the original RCT data by performing a {\em synthetic} RCT. 
Together, the two trials simulate the ideal, patient-level trial that is infeasible to implement in practice. 

% missingness mechanism 
\subsubsection*{Robustness to missingness mechanism}
A salient feature of the \SNN~estimator is that it is causally justified under a broad class of MNAR models. 
More specifically, \SNN~allows the missingness mechanism to be correlated with both (latent) patient-specific characteristics and (latent) factors that evolve over time, e.g., \SNN~permits correlation between Alice's decision to withdraw and her expected outcomes. 
This is important since patients often discontinue their assigned treatment due to adverse events, lack of tolerability, lack of efficacy, or simple inconvenience yet existing analytical methods are primarily designed for ``missing at random'' (MAR) settings \cite{dropout1}. 
MAR scenarios posit that the missingness pattern and counterfactual outcomes are conditionally independent given {\em observed} characteristics and outcomes. 
Because MAR imposes more stringent data generating assumptions than those compared to MNAR, the \SNN~estimator remains valid under MAR settings. 
%

% feasibility 
\subsubsection*{Feasibility \& efficacy} 
At this juncture, it is natural to ask whether \SNN~is even feasible to begin with. 
%A natural question at this juncture is whether \SNN~is feasible. 
%
At its core, \SNN~anchors on the underlying tensor having a low-dimensional structure. 
Intuitively, low-dimensionality implies that there is only a handful of canonical patient profiles and each patient's outcomes can be described by a unique weighted composition of outcomes associated with these canonical patients. 
\SNN~exploits this implication, which enables it to generate personalized predictions from relatively few patients (i.e., experiments). 
%under limited data constraints. 
%
We present data-driven diagnostics that check for the existence of synthetic patients and quantify the uncertainty in the estimates in the Materials and Methods section.
It is worth noting that low-dimensional datasets are empirically shown to be ubiquitous across numerous domains and are theoretically justified under many natural generative models \cite{Udell2017NiceLV, udell2018big, Chatterjee15, xu2017rates, pcr_jasa}. 
As such, low-dimensional tensors stand as the de-facto assumption in the tensor completion literature \cite{tensor_missing, gandy, anandkumar2014tensor, barak2016noisy}. 

%To demonstrate our approach, we utilize data from a 
We study the performance of \SNN~using Phase $3$ clinical trial conducted by TauRx Therapeutics on patients with Alzheimer's Disease (AD), a complex neurodegenerative disease with multiple cognitive and behavioral symptoms \cite{AD}. 
The data shows evidence of low-dimensional structure, which justifies and enables \SNN. 
Accordingly, we apply \SNN~towards the above two applications: (i) imputing unrecorded outcomes from dropouts given MNAR data and (ii) performing synthetic RCTs. 
In both studies, we provide predictions at the individual patient level and evaluate on a held-out test set of ground truth observations. 
Relative to several standard methods, we find that \SNN~performs well with average predictions errors roughly 7.8\% and 13.6\% lower than the next best estimator for the first and second applications, respectively. 
Our diagnostic results are also correlated with the prediction performance. 
%Our results suggest that the \SNN~estimator can be advantageous over standard estimators used to analyze trials and the diagnostic results are correlated with prediction performance. 
%
Altogether, %our studies showcase how 
our encouraging results provide evidence that 
\SNN~can not only be used to address an existing pain point in the current clinical trial workflow but also operate as a new tool towards the development of precision medicine. 

%% organization 
%\subsection*{Organization} 
%%
%Section~\ref{sec:results} formally describes our data and presents our empirical results. 
%%
%Section~\ref{sec:discussion} discusses generalizations of \SNN~in using observational data. 
%%
%Section~\ref{sec:mat_methods} describes the \SNN~estimator, as well as its operating assumptions and diagnostics. 
%%
%Section~\ref{sec:conclusion} summarizes our key takeaways. 

% RESULTS 
\section*{Results} \label{sec:results} 

% DATA
\subsection*{Data description} 
We consider Phase 3 clinical trial data conducted by TauRx Therapeutics. 
Data was collected across $N=1130$ patients with AD. 
The severity of AD is typically measured through cognitive exams such as the Alzheimer's Disease Assessment Scale-Cognitive Subscale (ADAS-Cog) \cite{adascog} or Mini Mental State Exam (MMSE) \cite{mmse}, both of which are continuous valued. 
%
%both scores are continuous valued with higher values corresponding to worse health. 
%
Before the trial's commencement, each patient's baseline ADAS-Cog and MMSE were recorded along their age and sex. 
We denote the pretreatment variables associated with patient $i$ as $X_i \in \Rb^4$. 
The trial itself spanned $T=5$ visits with $13$ weeks between each visit.
During the trial, patients were assigned to and observed under one of $A=3$ arms and their ADAS-Cogs were recorded during each visit. 
We remark that higher ADAS-Cog values indicate worsening health. 

\begin{figure}
\centering
\includegraphics[width=0.35\linewidth]{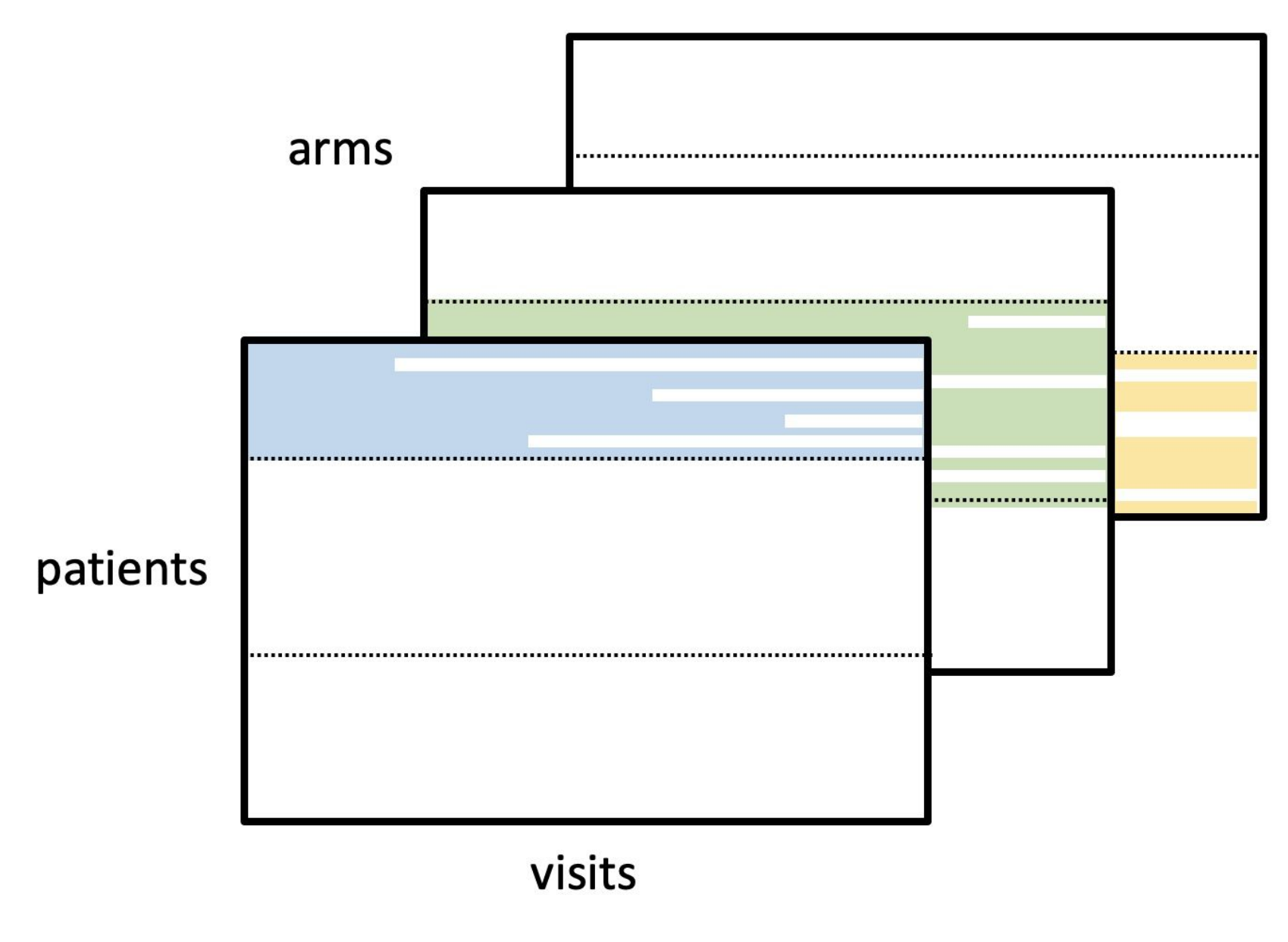}
\caption{Observed tensor of RCT outcomes, $\bY_\obs$, where rows index patients, columns index visits, and slices index treatment arms. Unobserved values are shown in white.}
\label{fig:tensor}
\end{figure}

We posit the existence of potential outcomes $Y_{it}(a) \in \Rb$ corresponding to the ADAS-Cog the $i$th patient would score during visit $t$ if they are assigned to arm $a$. 
We encode our observations into a $T \times N \times A$ tensor, $\bY_\obs = [Y_{ita}]$, where 
\begin{align} \label{eq:sutva} 
	Y_{ita} = \begin{cases}
		& Y_{it}(a), \text{ if } \Ac(i) = a \text{ and } D_{it} = 0
		\\ & \star, \text{ otherwise.}
	\end{cases}
\end{align}
Here, $\star$ represents a missing value, 
$\Ac(i)$ denotes patient $i$'s assigned arm, $D_{it}=1$ if patient $i$ drops out during visit $t$ and $D_{it}=0$ otherwise; we remark that $D_{it}=1$ implies $D_{i \tau}=1$ for all $\tau > t$. 
In words, \eqref{eq:sutva} states that patient $i$'s ADAS-Cog during visit $t$ can only be observed under their assigned treatment $a$ and if they did not drop out of the trial. 
See Figure \ref{fig:tensor} for a graphical depiction of the observed data pattern. 
Through this formulation, we define our objectives by defining the aspects of the tensor we wish to recover. 
For instance, using Figure \ref{fig:tensor}, the task of imputing dropout data translates to recovering the unobserved, white segments within each colored block. 
At the same time, estimating the counterfactual trajectory of outcomes for untested patient-treatment pairs corresponds to estimating the white blocks in each tensor slice that are entirely unobserved. 

% CASE STUDIES 
\subsection*{Empirical applications} 
We conduct two case studies to explore the performance of the \SNN~estimator. 
We remove dropouts within the clinical study to validate on a ground truth dataset. 
Given that standard trial designs calculate treatment effects based on final visits, we measure prediction accuracy based on the fifth (final) visit outcomes. 
For every arm $a$, we consider the (normalized) mean-squared-error (NMSE):  
\begin{align}
	\text{NMSE}(a) &:= \frac{ \sum_{i \in \Sc(a)} (Y_{i5} - \hEx[Y_{i5}(a)])^2}{\sum_{i \in \Sc(a)} Y_{i5}^2}.   \label{eq:mse}
\end{align} 
Here, $\Sc(a)$ denotes a patient subset of interest that will be formally defined in each experiment and $\hEx[Y_{i5}(a)] \in \Rb$ is an estimate of $\Ex[Y_{i5}(a)]$.  
The expectation with respect to $Y_{i5}(a)$ captures the inherent randomness in patient $i$'s outcome under treatment $a$. 
Note that \eqref{eq:mse} measures the average individual prediction performance. 

% DROPOUTS 
\subsubsection*{Imputing dropout data} 
We first study how \SNN~can impute unrecorded outcomes from dropouts. 
In practice, the reasons why patients withdraw can be hypothesized, but rarely (if ever) known. 
As such, we simulate dropouts under three missingness mechanisms in order to study \SNN's~robustness to the underlying data generating process: 
(i) patients withdraw uniformly at random, independent of all other factors; 
(ii) patients withdraw based on observed outcomes; 
(iii) patients withdraw based on observed and unobserved outcomes. 
In the literature, mechanism (i) is known as missing completely at random (MCAR), which is the most common assumption in the matrix and tensor completion literatures. 
Mechanisms (ii) and (iii), as discussed, correspond to MAR and MNAR, respectively.  
The MNAR model is arguably the most realistic as patients are more likely to withdraw if they are not responding well and do not expect to respond well to their assigned treatment moving forward. 
We examine the distribution of actual patient dropouts in the study and find that, within each arm, roughly $10\%$, $8\%$, $6\%$, and $4\%$ of patients withdraw during visits $2$, $3$, $4$, and $5$, respectively; 
for ease of notation, let $\rho(t)$ denote the percentage of dropouts for visit $t$, e.g., $\rho(2) = 0.1$. 
To simulate mechanisms (i)-(iii), we implement the following procedure: 
for every arm $a$ and visit $t \in \{2, \dots, 5\}$, 
\begin{enumerate}
	\item[(i)] MCAR: randomly sample $\rho(t) \cdot N_a$ patient dropouts, where $N_a$ denotes the size of arm $a$. Append the selected patients to the dropout set $\Sc(a)$. 
	
	\item[(ii)] MAR: for every patient $i$, define $b_i(t) = 1/(1+\exp(-\Delta(t)))$; here, $\Delta(t)$ is the exponentially weighted mean with span $t-1$ of $[Y_{i1} - Y_{i0}, \dots, Y_{it-1} - Y_{i0}]$, where $Y_{i0}$ is patient $i$'s baseline ADAS-Cog measured prior to the trial. Let $D_{it} = \text{Bernoulli}(b_i(t))$. Continue sampling dropouts until $\rho(t) \cdot N_a$ patients are reached and append the selected patients to the dropout set $\Sc(a)$.

	\item[(iii)] MNAR: for every patient $i$, define $b_i(t) = 1/(1+\exp(-\Delta(t)))$; here, $\Delta(t)$ is the exponentially weighted mean with span $5-(t-1)$ of $[Y_{it-1} - Y_{i0}, \dots, Y_{i5} - Y_{i0}]$. Let $D_{it} = \text{Bernoulli}(b_i(t))$. Continue sampling dropouts until $\rho(t) \cdot N_a$ patients are reached and append the selected patients to the dropout set $\Sc(a)$.

\end{enumerate} 
We interpret mechanisms (ii)-(iii). 
Under (ii), patient $i$'s dropout probability during visit $t$ is a function of their observed outcomes prior to visit $t$ with the most recent outcome during visit $t-1$ having the greatest influence on their decision. 
Similarly, under (iii), patient $i$'s dropout probability during visit $t$ is a function of their most recent outcome during visit $t-1$ and their future (unobserved) outcomes with the final outcome of the trial during visit $t=5$ having the greatest influence on their decision. 
Since higher ADAS-Cog values indicate worsening health, patients with more positive changes in ADAS-Cog from baseline (i.e., more positive values of $b_i(t)$) are more likely to withdraw in both scenarios.

We compare \SNN~against several baselines: (i) \Naive~estimator, (ii) last-observation-carried-forward (\LOCF), and (iii) \Matching.
For a dropout patient in arm $a$, the \Naive~estimator simply takes the average outcome of all patients within arm $a$ who stayed within the trial 
(compliers) as the counterfactual prediction. 
Intuitively, the \Naive~estimator requires high congruence between dropouts and compliers, thus it is a valid predictor if patients behave homogeneously. 
The \LOCF~predictor, commonly used in longitudinal studies \cite{missingdata1}, imputes missing data for each dropout by using their last available outcome. 
\Matching, as described in the Introduction, takes the average outcome of the patients most similar to the dropout patient of interest as its prediction. 

\begin{figure*}
\centering
\includegraphics[width=18.5cm]{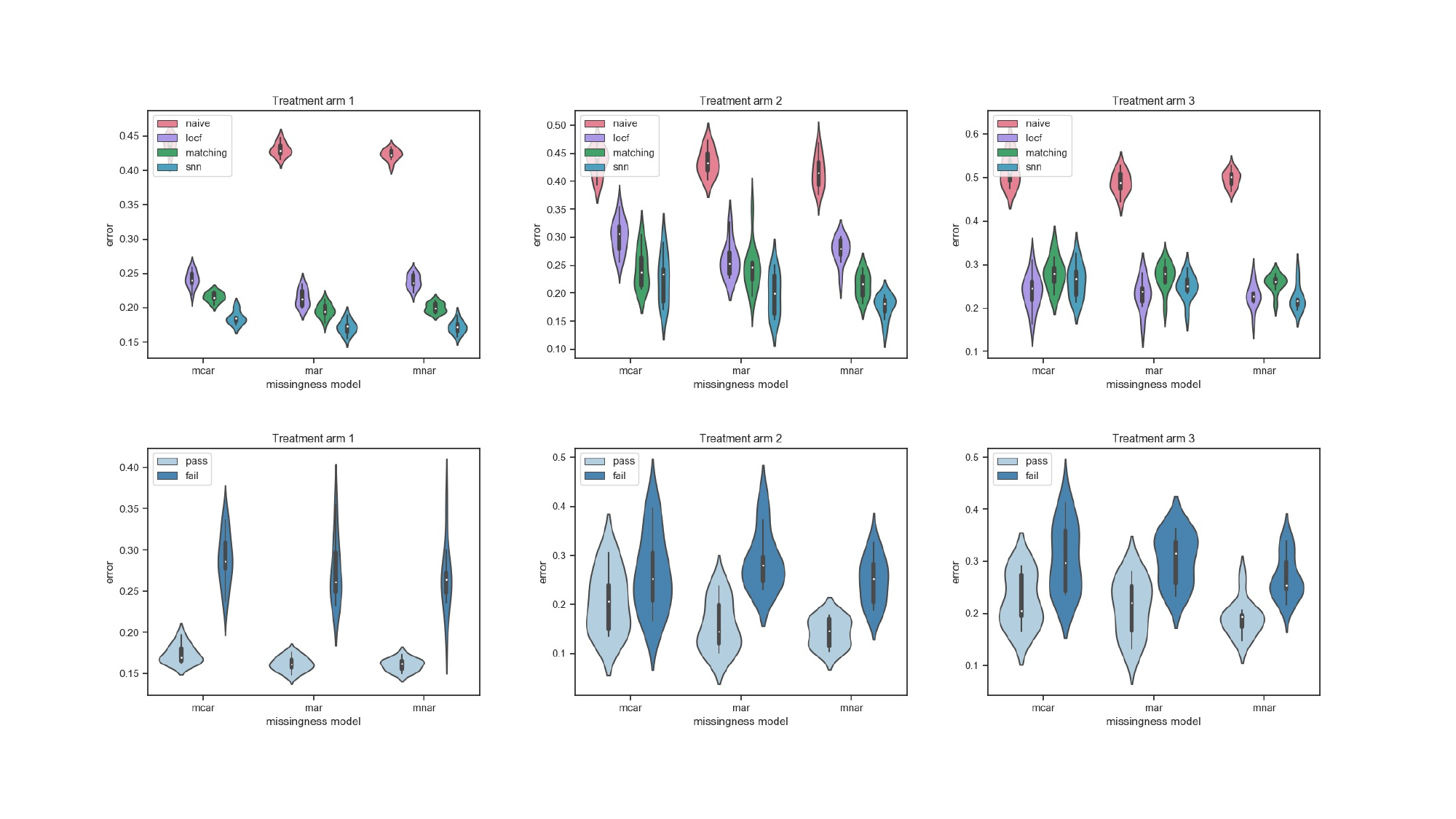}
\vspace{-40pt} 
\caption{Dropout simulation: (Top) Violin plots represent the range of NMSEs for every estimator across missingness models and treatment arms. We perform $10$ experimental repeats. The \Naive~estimator is shown in red, \LOCF~in purple, \Matching~in green, and \SNN~in blue. By and large, the \SNN~outperforms the baseline estimators. 
(Bottom) Violin plots represent the range of corresponding NMSEs for the \SNN~estimator when the data passes the diagnostics (light blue) and when it fails the diagnostics (dark blue) across missingness models and treatment arms. Consistently, the lower and higher errors correspond to passed and failed diagnostics, respectively. }
\label{fig:dropouts}
\end{figure*}

\begin{figure} [ht]
	\centering 
	\includegraphics[width=0.35\linewidth]{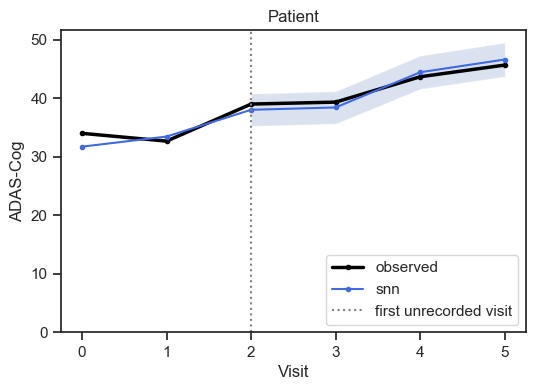}
	\caption{Comparison of the observed ADAS-Cog  (black) and that predicted by \SNN~(blue) for a simulated patient dropout. The shaded blue region represents the $95\%$ prediction uncertainty band. The vertical line represents the first unrecorded visit by the patient. Visit $0$ represents the baseline ADAS-Cog measured prior to the trial. The \SNN~synthetic model is learned using data to the left of the vertical line while the data to the right of the vertical line is the held-out ground truth used for validation.} 
	\label{fig:dropout} 
\end{figure}

We present our summarized results across 10 experimental repeats in the top set of plots in Figure~\ref{fig:dropouts}; an example patient simulation is provided in Figure~\ref{fig:dropout} for \SNN~only. 
In every iteration, each patient dropout's remaining trajectory is estimated using data from his or her outcomes prior to the withdrawal as well as the observations across compliers. 
Our validation is thus conducted on the held-out trajectories across patient dropouts, which are not utilized to learn any of the models. 
As Figure~\ref{fig:dropouts} suggests, \SNN, by and large, outperforms the baselines across missingness models and treatment arms with an average prediction error of roughly 7.8\% lower than the next best approach. 

% SYNTHETIC RCT
\subsubsection*{Running synthetic RCTs} 
Next, we explore how \SNN~can leverage standard RCT data to perform synthetic RCTs. 
Ideally, to carry out this experiment, we would use our raw RCT data, where each patient was assigned to and observed under a single arm, to predict the outcomes for all other untested patient-treatment combinations and evaluate on these untested pairs.
Of course, these outcomes are counterfactual, which renders their evaluation to be impossible. 
As such, we will emulate aspects of the ideal experiment by withholding half of our observed data and attempting to recreate said outcomes instead. 
More specifically, for every arm $a$, we randomly select half of the patients as a test set, $\Sc(a)$, and withhold their RCT outcomes. 
We define the remaining patients as our train set and retain their RCT outcomes. 
In both patient sets, we observe the pretreatment variables $X_i$. 
Our interest is to recreate the RCT outcomes of the test patients using the RCT outcomes of the train patients. 
Under this setup, the retained observations act as our raw RCT data while the withheld observations serve as our untested pairs. 

\begin{figure*}[ht]
\centering
\includegraphics[width=12cm]{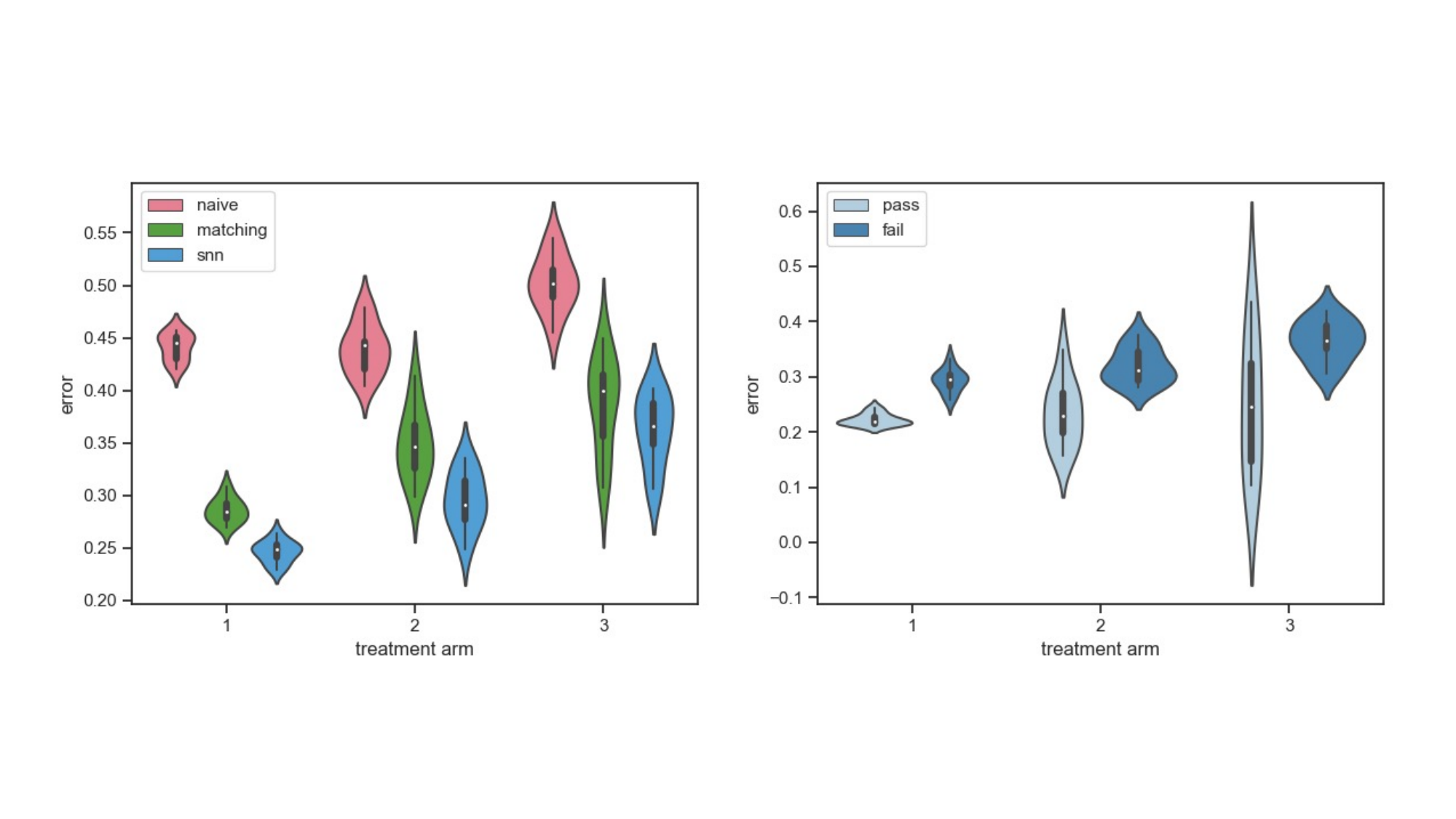} 
\vspace{-30pt} 
\caption{Synthetic RCT simulation: (Left) Violin plots represent the range of NMSEs for every estimator across treatment arms. We perform $10$ experimental repeats. The \Naive~estimator is shown in red, \Matching~in green, and \SNN~in blue. Throughout, the \SNN~outperforms the baseline estimators. 
(Right) Violin plots represent the range of corresponding NMSEs for the \SNN~estimator when the data passes the diagnostics (light blue) and when it fails the diagnostics (dark blue) across treatment arms. Consistently, the lower and higher errors correspond to passed and failed diagnostics, respectively.}
\label{fig:synthRCT}
\end{figure*}

We compare \SNN~against two baselines from the previous dropout study: (i) \Naive~estimator and (ii) \Matching. 
We reemphasize that the models in this experiment are strictly built on pretreatment variables while the dropout experiment models were given access to both pretreatment variables and RCT outcomes prior to a patient's withdrawal. 
As such, we remove \LOCF~as a baseline since we do not have any clinical outcomes to extrapolate from in constructing our counterfactual predictions. 
We summarize our results across 10 experimental repeats in Figure~\ref{fig:synthRCT}. 
Our empirical findings show that \SNN~outperforms both baselines across all treatment arms with an average prediction error of roughly 13.6\% lower than the next best method. 
However, it is worth noting that the errors induced in this setting are naturally larger than those in our dropout study. 
This is likely due to the models in this experiment accessing less data (pretreatment variables) compared to the dropout experiment. 
%the fact that models here are strictly built on pretreatment variables while the dropout study models were given access to both pretreatment variables and RCT outcomes prior to a patient's withdrawal. 
%
As such, utilizing a richer set of clinical outcomes collected prior to the trial, e.g., outcomes associated with a patient's natural history of disease, should improve statistical models such as \SNN. 

\subsection*{Diagnostics}
As with any causal inference method, it is critical to examine the suitability of the method for the chosen application and its counterfactual predictions. 
Ignoring these safeguards risks drawing specious conclusions. 
To avoid these pitfalls, we inspect structural properties of the data and perform a series of diagnostics (i.e., model checks). 

\begin{figure}[!t]
	\centering 
	\includegraphics[width=0.35\linewidth]{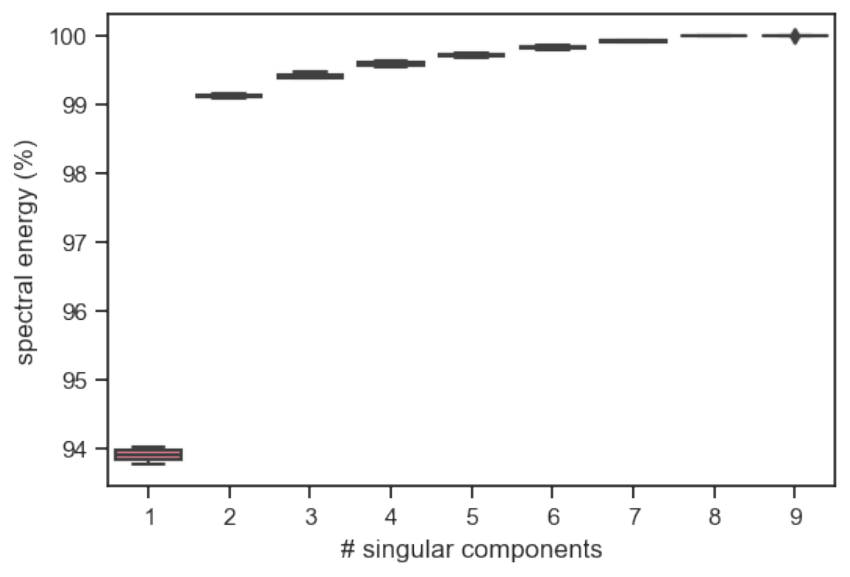}
	\caption{Spectral energy captured by the top $9$ singular values. The box-and-whiskers represent the range of values over all treatment arms. Over $99\%$ of the spectral energy is captured by the top two singular components in each arm, which suggests that the low-rank assumption is reasonable.} 
	\label{fig:spectral} 
\end{figure}

We first plot the spectral profile for each arm in Figure~\ref{fig:spectral}. 
For every arm $a$, we construct the concatenated matrix $[\bX_a, \bY_a]$ of pretreatment variables and RCT outcomes, respectively, where the rows of $\bX_a$ and $\bY_a$ correspond to the patients in arm $a$ for which all RCT outcomes are observed. 
These are the same patients used in our empirical studies (i.e., actual dropouts are not included in our spectral analysis). 
As Figure~\ref{fig:spectral} shows, over $99\%$ of the spectral energy is captured by the top two singular values in every arm $a$, which indicates that each concatenated matrix is low-rank. 
This provides evidence that the underlying tensor of potential outcomes is also low-rank, which suggests the suitability of \SNN~for our studies. 
Intuitively, these results indicate that, within each arm, there are effectively two canonical patient profile types and every patient can be well-described by these two types. 

Next, we perform diagnostics to examine the quality of \SNN's~counterfactual predictions; the details of the diagnostics are provided in the Methods section. %the Material and Methods section. 
The diagnostics consist of computing two statistics whose values range from $0$ to $1$ and are conducted for every $(i,t,a)$ tuple. 
At a high level, the purpose of the first statistic is to test (to the extent possible) the existence of a synthetic patient $i$ formed from patients in group $a$. 
The second statistic is designed to test if the synthetic model learned from pretreatment variables and RCT outcomes prior to time $t$ can generalize to yield a reliable estimate for visit $t$ under treatment $a$. 
The values of the statistics can be interpreted as the percentages of the data that {\em cannot} be explained by the prediction model. 
%
%For both statistics, 
Accordingly, lower values are more desirable but higher values are more informative. 
That is, both statistics should be interpreted as one-sided tests: if either statistic is close to $1$, then the counterfactual prediction for the $(i,t,a)$th tuple is likely to be poor; 
even if both values are close to $0$, however, perfect prediction is still not guaranteed.  
The notions of ``low'' and ``high'' values are at the discretion of the researcher. 
We define low and high as values less than and greater than $\alpha = 0.2$, respectively. 
The diagnostic results for both studies are visualized in the bottom plots of Figures~\ref{fig:dropouts} and \ref{fig:synthRCT}. 
We say the model passes the diagnostics if both statistics are low and fails otherwise. 
Across all experimental setups, models that pass the diagnostics yield much smaller errors than those that fail the diagnostics---roughly 51.3\% and 40\% lower average prediction errors in the first and second studies, respectively. 
These findings suggest the the diagnostics help flag fragile models and are largely correlated to the quality of \SNN's~estimates. 

% DISCUSSION 
\section*{Discussion} \label{sec:discussion} 
We discuss the application of \SNN~to clinical trials and beyond. 

% EXPERIMENTAL DATA
\subsection*{Applications with RCT data}
\SNN~enables researchers to extract more data and more insights from their trials without having to perform additional experiments. 
In particular, our studies suggest that \SNN~can be used for two important applications: data completion and data augmentation. 

Our first study provides evidence that \SNN~can help power and de-bias a RCT by 
%can convert an under-powered, biased RCT into a fully-powered, bias-corrected RCT. 
%
%This is achieved by 
imputing unrecorded outcomes associated with dropouts to simulate the counterfactual scenario where all patients are compliant. 
As a byproduct, researchers are better equipped to estimate the true treatment effects. 
%By doing so, researchers are then able to draw inferences from a complete dataset to better measure the true treatment effects.  

Our second study indicates that \SNN~can leverage standard RCT data to run synthetic RCTs.
Combined, the physical and virtual trials emulate the ideal, patient-level RCT, where each patient is observed under every treatment. 
Having access to an augmented labeled dataset can then help researchers identify potential subpopulations of responders and non-responders for every treatment. 
This can aid researchers in making downstream decisions. 
More specifically, researchers may wish to further investigate subpopulation-treatment pairs that yield promising outcomes and abandon pairs that yield undesirable outcomes. 
Examples include the modification of inclusion-exclusion criteria and early termination. 
%One instance of said scenario is to modify inclusion-exclusion criteria in adaptive trials. 
%
Therefore, \SNN~can serve as a recommendation engine to strategically decide the experiments to prioritize and a tool towards the development of precision medicine. 

Collectively, the two applications provide researchers with a better view of their historical and current trials and a clearer direction for future trials. 

% OBSERVATIONAL DATA
\subsection*{Applications with observational data}  
\begin{figure*}
	\centering 
	\includegraphics[width=12cm]{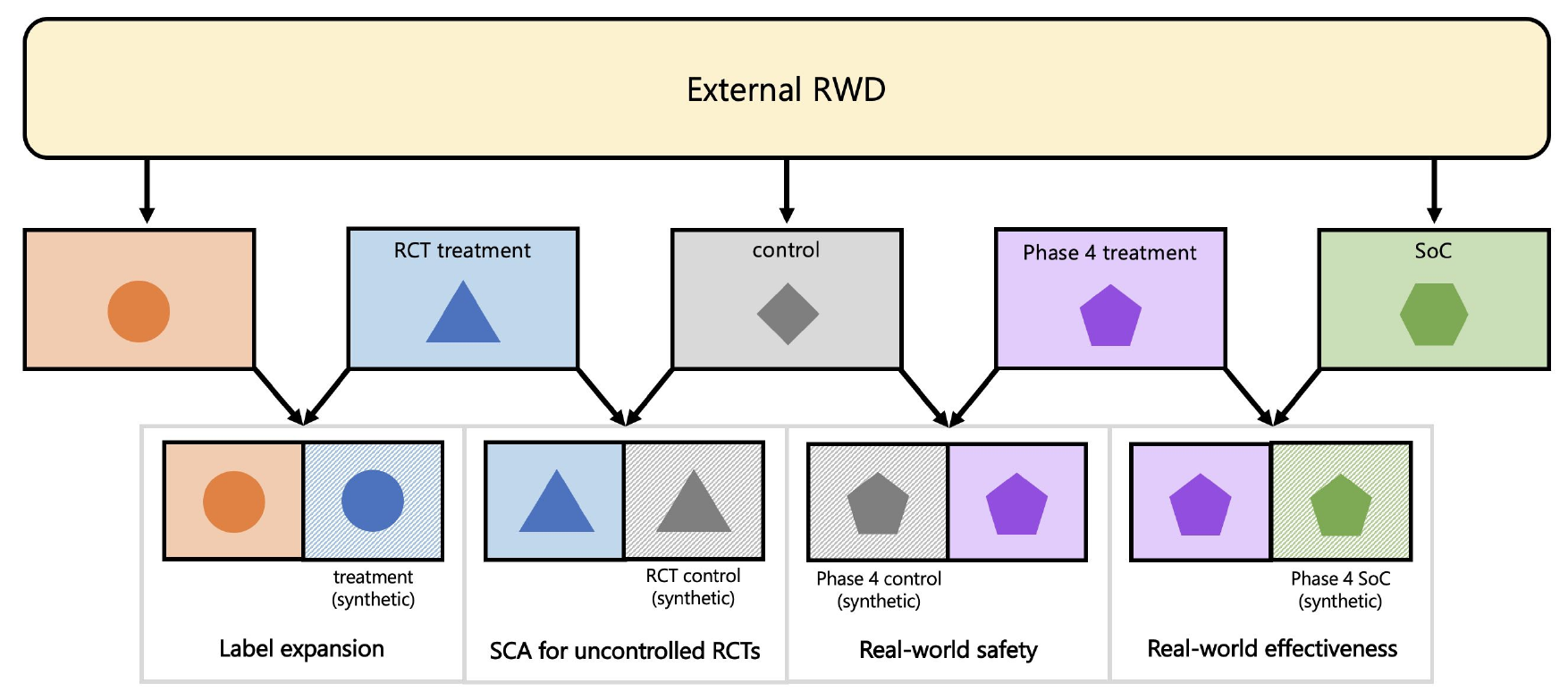}
	\caption{Potential inputs and outputs for four applications of extracting RWE from RWD. 
	Colors correspond to therapies and shapes correspond to patient cohorts. 
	The solid colored rectangles are observed and dashed colored rectangles are synthetic. 
	%To facilitate our discussion, we suppose a treatment is designed for patients within subpopulation A. 
	In label expansion, \SNN~constructs a synthetic cohort under treatment from RCT patients in the treatment arm. In SCA for uncontrolled trials, \SNN~constructs a SCA from real-world control patients. In demonstrating real-world safety, \SNN~constructs a synthetic cohort of Phase 4 patients under control from real-world control patients. In demonstrating real-world effectiveness, \SNN~constructs a synthetic cohort of Phase 4 patients taking the SoC.} 
	\label{fig:rwd} 
\end{figure*}

We devote this section to discussing how \SNN~can utilize real-world data (RWD) to lay the foundations for real-world evidence (RWE). 
The U.S. Food and Drug Administration (FDA) defines RWD as observational data collected in real-world settings and RWE as medical insights derived from analyzing RWD. 
RWD is extracted from numerous sources spanning electronic medical records, health insurance claims, and patient-reported outcomes. 
Below, we discuss four applications of \SNN~in this context and summarize the input-output relations for each in Figure~\ref{fig:rwd}. 

% SCA
\subsubsection*{Synthetic control arms} 
The first application is in regards to uncontrolled, single-arm trials, which are increasingly common in rare and orphan disease areas. 
These disease areas have small patient populations with possibly life-threatening conditions and often lack established standards-of-care (SoC). 
This translates to clinical trials with insufficient statistical power and raises ethical concerns on administering placebo. 
To tackle these issues, the U.S. FDA and European Medicines Agency (EMA) have taken several initiatives to allow for novel design approaches that use external control data \cite{us2019submitting, committee2006guideline}. 
Many standard causal inference methods have been employed to construct synthetic control arms from RWD, including the \Naive~and \Matching~methods used in our case studies \cite{sc_arms}. 
Given our encouraging empirical findings, we believe that \SNN~can also be used to evaluate the comparative effectiveness of a treatment using external control data. 
In particular, \SNN~can construct a synthetic twin of every in-trial patient under control using a weighted combination of external control patients. 
The cohort of synthetic twins then forms the synthetic control arm (SCA).  
%

% RWE Safety & Effectiveness 
\subsubsection*{Real-world safety \& effectiveness} 
In most countries, to provide evidence of safety, pharmaceutical companies must perform a comprehensive surveillance of their newly-approved treatments (pharmacovigilance) to identify potential hazards that may be harmful to their patients. 
In particular, pharmaceutical companies are required to conduct adverse event reporting. 
One primary purpose of these reports is to determine whether adverse events are caused by their treatments. 
Determining causality, however, is an arduous undertaking due to the lack of reliable data. 
Current assessments involve observational studies where the incidence of adverse events is compared between a patient population taking the treatment and a natural control group. 
Again, \SNN~can be employed to construct an artificial control group. 
That is, for every patient under treatment, \SNN~constructs a synthetic twin under control from real-world control patients. 
If the incidence of an adverse event is statistically significantly higher in the observed treatment group versus the synthetic control group, then there is evidence that the treatment contributes to the adverse event. 
To demonstrate effectiveness, companies must provide evidence that their newly-released treatment is competitive with the established SoC. 
Towards this, \SNN~can be used to construct synthetic cohorts under SoC. 
Similar to the case of SCA, \SNN~generates a synthetic twin under SoC for every patient under the new treatment. 
Comparing the outcomes between the observed and artificial groups allows researchers to assess the comparative effectiveness of the new treatment versus SoC. 
%
%If statistically significantly more favorable outcome, then possibly strengthen evidence for differentiation 

% LABEL EXPANSION
\subsubsection*{Label expansion} 
Finally, we discuss how \SNN~can play a role in label expansion. 
In this setting, companies seek RWE to showcase performance of their treatment under off-label indications. 
To discuss \SNN's utility in this setting, consider a cohort of off-label patients and a cohort of on-label patients taking the newly-released treatment. 
We define off-label patients as individuals that are underrepresented or not represented in the RCT, e.g., these patients may satisfy the RCT inclusion-exclusion criteria but belong to a different ethnic group than that of the RCT participants or exhibit similar covariates as the on-label patients but do not strictly satisfy the inclusion-exclusion criteria.
For every off-label patient, \SNN~constructs a synthetic twin under the newly-released treatment as a weighted combination of on-label patients. 
%
%If the on-label patient data was collected from a RCT, then 
This process effectively simulates a synthetic trial of off-label patients under said treatment. 
The outcomes of the simulation, therefore, provide researchers with evidence in support of or against expanding the treatment to the off-label patient subpopulation. 

% validity 
\subsection*{Validity for real-world studies}
In all of the applications above, \SNN~uses RWD to construct a synthetic comparison group. 
Since RWD lacks randomization, there will naturally be a healthy degree of skepticism at the robustness of \SNN~to confounded data. 
We take this opportunity to reemphasize that \SNN~is justified under a large class of natural MNAR settings, e.g., patients are allowed to seek (or avoid) treatment based on their unique characteristics and expected health outcomes; we formalize this claim in the Materials and Methods section.  
We also argue that \SNN, like matching, offers interpretability and transparency, which can be critical for real-world studies since regulatory bodies may be involved. 
This is in stark contrast to black-box technologies based on deep learning and ensemble methods (e.g., random forests) that promise strong prediction capabilities but do not guarantee interpretable findings \cite{sc_arms}. 

However, \SNN~is not immune to other challenges that arise from RWD. 
For instance, treatments may be administered differently in real-world settings from RCTs and outcomes are often sampled at irregular time points. 
It is critical to exercise a considerable amount of care in devising proper implementation strategies to account for these nuances.

\section*{Method} \label{sec:mat_methods} 
%
%In this section, we formally present our causal framework and the \SNN~estimator. 
%

% framework 
\subsection*{Causal framework}
To discuss causality, we use the language of potential outcomes \cite{neyman, rubin}. 
We remark that \eqref{eq:sutva} is known in the causal inference literature as the stable unit treatment value assumption (SUTVA). 
SUTVA looks innocuous but packs in several important nuances.
Namely, SUTVA implies that the observation of one patient is unaffected by the treatment of other patients (i.e., no spillover effects) and that each patient in arm $a$ is treated identically (i.e., there is only one form of treatment $a$) \cite{imbens_rubin_2015}. 
SUTVA is more widely accepted in RCTs, but is naturally more difficult to verify in RWD; therefore, as noted in our Discussion section, RWD and RWE applications require greater precautions to take into account any violations of SUTVA. 

First, we organize our potential outcomes into a $T \times N \times A$ tensor, $\bY = [Y_{it}(a)]$. 
Our tensor formulation is critical. 
On the surface, it provides a convenient representation of our data and potential outcomes. 
Underneath and more importantly, the structural properties of the tensor---which can be empirically inspected and validated (to an extent)---allow us to properly justify the \SNN~approach and establish the class of permissible data generating processes. 
Towards these discussion points, we posit that $\bY$ follows a low rank tensor factor model: 
\begin{align} \label{eq:po} 
	Y_{it}(a) &= \langle u_i, v_t(a) \rangle + \varepsilon_{it}(a), 
\end{align} 
where $u_i \in \Rb^r$  is a latent factor specific to patient $i$, $v_t(a) \in \Rb^r$  is a latent factor specific to visit $t$ and arm $a$, $\varepsilon_{it}(a) \in \Rb$ is a mean zero random variable that models the stochasticity in ADAS-Cog outcomes, and $r \ll \min\{N, T \times A\}$. 
Notably, \eqref{eq:po} is implied by a tensor with low canonical polyadic (CP) rank, which is the standard assumption within the tensor completion literature. 
We also posit 
\begin{align} \label{eq:covariate} 
	X_{i\ell} = \langle u_i, w_\ell \rangle + \eta_{i\ell},
\end{align} 
where $u_i$ is defined as above, $w_\ell \in \Rb^r$ is a latent factor specific to pretreatment variable (covariate) $\ell$, $\eta_{i \ell} \in \Rb$ is a mean zero residual that models measurement noise, and $r \ll d$. 
That is, $\bX = [X_i] \in \Rb^{N \times d}$, is a low-rank matrix of patient covariates. 
Together, \eqref{eq:po} and \eqref{eq:covariate} state that the latent potential clinical outcomes and observed pretreatment variables are connected by the same latent patient factors, which span a $r$-dimensional subspace of $\Rb^N$ typically with $r \ll N$.
Hence, even if the number of patients $N$ is large, there are only $r$ canonical patient profiles among them and each patient can be expressed as a linear combination of these $r$ profiles with high probability (w.h.p.). 
To see this, consider imputing the $(i,t,a)$th expected potential outcome and let $\Pc \subseteq \{j: \Omega_{jta} = 1\}$. 
Since $u_i \in \Rb^r$, if $\text{span}(\{u_j: j \in \Pc\}) = \Rb^r$, then $u_i \in \text{span}(\{u_j: j \in \Pc \})$. 
More generally, if the latent patient factors are randomly sampled from a sub-Gaussian distribution, then $\text{span}(\{u_j: j \in \Pc\}) = \Rb^r$ w.h.p., provided $| \Pc | \ge r$ is chosen to be sufficiently large \cite{vershynin2018high}. 

To formalize the class of permissible data generating processes (i.e., MNAR models), we define $\bOmega \in \{0,1\}^{N \times T \times A}$ as the tensor whose elements indicate the $(i,t,a)$ tuples that are observed within $\bY$, i.e., $\Omega_{ita} = 1$ if $\Ac(i) = a$ and $D_{it} = 0$, and $\Omega_{ita} = 0$ otherwise. 
For every $(i,t,a)$ tuple, we assume $\bOmega \independent \varepsilon_{it}(a) | \LFc$, where 
$$\LFc = \{u_i, v_t(a), w_\ell : i \in [N], t \in [T], a \in [A], \ell \in [d]\}$$
represents the collection of latent factors.\footnote{Let $[M] = \{1, \dots, M\}$ for any positive integer $M$.} 
In our context, the latent factors determine not only a patient's treatment assignment and decision to withdraw, but also the patient's clinical outcomes and covariates. 
Combined with \eqref{eq:po}, $\bOmega \independent \varepsilon_{it}(a) | \LFc$ implies that $\bOmega \independent Y_{it}(a) | \LFc$, i.e., conditioned on the latent factors, the potential outcomes are independent of the treatment assignment and decision to withdraw.
We also assume $\bOmega \independent \eta_i | \LFc$, where $\eta_i = [\eta_{i\ell}] \in \Rb^d$. 
Combined with \eqref{eq:covariate}, this implies that $\bOmega \independent \bX | \LFc$.
All in all, the probability of observing an outcome for the $(i,t,a)$th tuple can be influenced by latent characteristics specific to patient $i$, visit $t$, arm $a$, and pretreatment variables $\ell \in [d]$, which themselves can be arbitrarily correlated. 
In other words, the latent factors can also be latent confounders. 

Our assumptions yield the following identification result: 
for every $(i,t,a)$ tuple and set $\Pc \subseteq \{j: \Omega_{jta} = 1\}$ with $| \Pc | \ge r$, there exists a vector of linear weights, $\beta \in \Rb^{|\Pc|}$, such that 
\begin{align} \label{eq:identification}
	\Ex[Y_{it}(a) ~|~ u_i, v_t(a)] &= \sum_{j \in \Pc} \beta_j \Ex[Y_{jta} ~|~ \bOmega, \LFc]. 
\end{align} 
In words, \eqref{eq:identification} states that patient $i$'s expected potential outcome during visit $t$ under treatment $a$ can be written as a linear combination, defined by $\beta$, of expected observed outcomes associated with those patients in $\Pc$. 
Importantly, \eqref{eq:identification} does not require patient $i$ to be assigned to arm $a$---this is precisely what enables the construction of a synthetic patient $i$ under different treatments. 
Additionally, \eqref{eq:identification} suggests that $\beta$ is the key object to estimate. 
Towards this, we note that our assumptions also establish that for every patient $i$
\begin{align} \label{eq:covariate_identification} 
	\Ex[X_i ~|~ u_i, w_1, \dots, w_d] = \sum_{j \in \Pc} \beta_j \Ex[X_j ~|~ \bOmega, \LFc]. 
\end{align}  
This implies that $\beta$ can be learned from both treatment and pretreatment variables. 
In what follows, we describe the \SNN~estimator, which provides one procedure that leverages both variable types to learn the unique $\beta$ for every $(i,t,a)$ tuple of interest. 
%

% methodology 
\subsection*{\SNN~estimator} 
Without loss of generality, consider estimating the expected potential outcome for the $(i,t,a)$th tuple. 
Let $\Pc = \{j: \Omega_{jta}=1\}$ denote the set of patients in arm $a$ whose ADAS-Cogs are recorded during visit $t$. 
Recall that $\Omega_{jta}=1$ implies that $\Omega_{j\tau a}=1$ for all $\tau < t$. 
Further, let $\Tc = \{\tau < t: \Omega_{i \tau a} = 1\}$ denote the set of visits for which patient $i$'s outcomes are recorded. 
If patient $i$ withdrew before the start of the trial or was assigned to a different arm $a' \neq a$, then $\Tc = \emptyset$. 

Let $Z_{i \Tc} = [X_i, Y_{i \Tc}]$, where $Y_{i \Tc} = [Y_{i \tau a}: \tau \in \Tc]$. 
Further, let $K \ge 1$ be a user-specified hyper-parameter that defines the synthetic neighborhood size. 
Given $K$, we randomly partition $\Pc$ into $K$ subgroups, denoted as $\Pc_k$,  of roughly equal size, i.e., $|\Pc_k| \sim |\Pc| / K$.
For every $k \in [K]$, let
$\bY_{\Pc_k \Tc} = [Y_{j \tau a}: j \in \Pc_k, \tau \in \Tc]$ 
and 
$\bX_{\Pc_k} = [X_{j}: j \in \Pc_k]$ denote the matrices of clinical outcomes and covariates, respectively, for the $k$th subgroup in arm $a$.  
Let $\bZ_{\Pc_k \Tc} = [\bX_{\Pc_k}, \bY_{\Pc_k \Tc}]$ denote the concatenated matrix. 
We write its singular value decomposition (SVD) as 
%
%\begin{align}
	$\bZ_{\Pc_k \Tc} = \sum_{\ell \ge 1} \hsigma_\ell \hu_\ell \hv_\ell^\top$,
%\end{align} 
%
where $\hsigma_\ell$ are the singular values, and $\hu_\ell, \hv_\ell$ are the left and right singular vectors, respectively. 
For any integer $b \le \min\{|\Pc_k|, d+|\Tc|\}$, let $\bhU_b = [\hu_1, \dots, \hu_b] $ denote the matrix formed by the top $b$ left singular vectors of $\bZ_k$. 
We define $\bhV_b = [\hv_1, \dots, \hv_b]$ and $\bhSigma_b = \text{diag}(\hsigma_1, \dots, \hsigma_b)$ similarly. 

\begin{enumerate}
	\item[(a)] {\em Model learning}: for every $k \in [K]$, compute
	\begin{align} \label{eq:pcr} 
		\hbeta^{(k)} &= \argmin_{\beta \in \Rb^{|\Pc_k|}} 
		\| Z_{i \Tc} - \bZ^\top_{\Pc_k \Tc} \beta \|_2^2 
		~\text{ subject to } (\bI - \bhU_{b_k} \bhU_{b_k}^\top) \beta = 0.~
	\end{align}
	
	\item[(b)] {\em Diagnostics}: for every $k \in [K]$, compute
	\begin{align} 
		\theta_k &= \frac{ \| (\bI - \bhV_{b_k} \bhV_{b_k}^\top) Z_{i \Tc} \|_2}{ \| Z_{i \Tc} \|_2},
		\quad \phi_k = \frac{ \| (\bI - \bhU_{b_k} \bhU_{b_k}^\top) Y_{\Pc_k t} \|_2} {\| Y_{\Pc_k t} \|_2}, \label{eq:diagnostics} 
	\end{align} 
	where $Y_{\Pc_k t} = [Y_{jta}: j \in \Pc_k]$. 
	For a pre-specified tolerance $\alpha \in (0,1)$, discard the $k$th model if $\theta_k \ge \alpha$ or $\phi_k \ge \alpha$. 
	Define $\Kc = \{k: \theta_k, \phi_k < \alpha\}$ as the set of retained models. 
	
	\item[(c)] {\em Counterfactual prediction}: 
	\begin{align} \label{eq:prediction} 
		\hEx[Y_{it}(a)] &= \frac{1}{|\Kc|} \sum_{k \in \Kc} \hEx[Y^{(k)}_{it}(a)],
		~\text{ where } \hEx[Y^{(k)}_{it}(a)] =  \langle Y_{\Pc_k t}, \hbeta^{(k)} \rangle. 
	\end{align} 
	%
	%where $\hEx[Y^{(k)}_{it}(a)] =  \langle Y_{\Pc_k t}, \hbeta^{(k)} \rangle$.
	
	\item[(d)] {\em Inference}: 
	the middle $95\%$ of our estimates $\hEx[Y^{(k)}_{it}(a)]$ across $k \in \Kc$ forms a $95\%$ prediction interval. 
	If $|\Kc|=1$, then we can use the following formula: 
	\begin{align}
		&\Ex[Y_{it}(a)] \in \Big[ \hEx[Y_{it}(a)] \pm z_{\text{ci}} \cdot  \hnu \cdot \sqrt{ \big\langle Y_{\Pc t}, \bhV_{b} \bhSigma_{b}^{-2} \bhV_{b}^\top Y_{\Pc t} \big\rangle} \Big], 
		\\
		&Y_{it}(a) \in \Big[ \hEx[Y_{it}(a)] \pm z_{\text{ci}} \cdot  \hnu \cdot  \sqrt{1 + \big\langle Y_{\Pc t}, \bhV_{b} \bhSigma_{b}^{-2} \bhV_{b}^\top Y_{\Pc t} \big\rangle} \Big],
	\end{align} 
	where $z_{\text{ci}}$ is the $z$-interval that defines the confidence interval range (e.g., $z_{\text{ci}}=1.96$ for a $95\%$ confidence interval) and  
	$\hnu$ is an estimate of the standard deviation of the stochastic terms in the potential outcomes. We remark that a standard estimate of $\hnu = \theta \| Z_{i \Tc} \|_2 /(|\Tc|+d)$, where $\theta$ is defined in \eqref{eq:diagnostics}. 
\end{enumerate}
{\em Interpretation.} 
To begin, we note that \eqref{eq:pcr} is known in the statistics literature as principal component regression (\PCR). 
\PCR~is a form of regularized regression that enforces spectral sparsity as it only retains the top $b_k$ singular values to learn a linear model. 
Such an approach is well justified when $\bZ_k$ exhibits low-rank structure. 
Notably, \eqref{eq:pcr} admits a simple closed form solution: 
%\begin{align}
	$\hbeta^{(k)} = ( \sum_{\ell=1}^{b_k} (1/\hsigma_\ell) \hu_\ell \hv_\ell^\top ) Z_{i \Tc}$. 
%\end{align} 
%
There are numerous approaches to choose $b_k$; 
we recommend the universal thresholding approach of \cite{donoho14}. 

Next, we discuss the diagnostic values $\theta_k, \phi_k$. 
To this end, we underscore that faithful recovery of the $(i,t,a)$th expected potential outcome requires that the row and column spaces of $\bZ_{\Pc_k \Tc}$ are sufficiently ``rich''. 
At a high level, the row space---generated by $\Pc_k$---determines the existence of a synthetic patient $i$ formed by the subset of patients in $\Pc_k$, while the column space---generated by the covariates and $\Tc$---determines \SNN's~ability to learn this synthetic model. 
More formally, the row space determines the validity of \eqref{eq:identification} while the column space determines if the learned model can generalize to the data point $Y_{\Pc_k t}$. 
With these concepts in mind, we remark that $\theta_k$ and $\phi_k$ are designed to test the former and latter conditions, respectively, to the extent possible. 
In particular, $\theta_k$ measures the $\ell_2$-distance of patient $i$'s covariates and outcomes, $Z_{i \Tc}$, from the row space of $\bZ_{\Pc_k \Tc}$;\footnote{We highlight that $\theta_k$ is equivalent to the normalized training root mean-squared-error, i.e., $\theta_k = (1/ \| Z_{i \Tc} \|_2) \cdot \| Z_{i \Tc} - \bZ^\top_{\Pc_k \Tc} \hbeta^{(k)} \|_2 $.} 
$\phi_k$ measures the $\ell_2$-distance of the outcomes associated with patients in $\Pc_k$ during visit $t$, $Y_{\Pc_k t}$, from the column space of $\bZ_{\Pc_k \Tc}$. 
By construction, both $\theta_k, \phi_k \in [0,1]$, and can be interpreted as one-sided tests. 
That is, the greater the dimension of the row and column spaces (i.e., the rank of $\bZ_{\Pc_k \Tc}$), the greater the probability that $\theta_k, \phi_k \approx 0$. 
However, values close to $0$ do not guarantee that \eqref{eq:prediction} will be near perfect; values close to $1$, however, should raise extreme caution about the prediction quality of \eqref{eq:prediction}. 
Therefore, a researcher should find $\theta_k, \phi_k$ to be useful measures in discarding poor counterfactual predictions but not in certifying high-quality counterfactual predictions.   

Finally, the inference calculations for $|\Kc|=1$ are derived from the standard linear regression confidence and prediction interval equations using the plug-in estimators for the latent left singular vectors and singular values of $\Ex[\bZ_{\Pc \Tc}]$. 
These derivations assume homoskedasticity. 
For heteroskedastic intervals, see \cite{shen2022root}. 

\subsection*{Guidelines \& Regulations}
All methods were carried out in accordance with relevant guidelines and regulations. 
All experimental protocols were approved by the Copernicus IRB group and Western Institutional Review Board (WIRB). 
Informed consent was obtained from all subjects and/or their legal guardian(s). 
 
%}
%\showmatmethods{} % Display the Materials and Methods section
 
% CONCLUSION 
\section*{Conclusion} \label{sec:conclusion} 
This work utilizes the recently proposed \SNN~estimator to provide personalized predictions from population-level RCTs. 
Using Phase 3 RCT data on patients with Alzheimer's Disease, we examine \SNN~for two applications: 
imputing dropout data and running synthetic RCTs. 
Relative to several current methods, we find that \SNN~performs well in these contexts. 
%Our results provide evidence of \SNN's~validity in these contexts and its benefits over current estimators. 
%
Our encouraging results demonstrate that \SNN~can help power and de-bias RCTs, and can also operate as a recommendation engine to strategically prioritize experiments and thus advance the initiative of precision medicine. 
%examine how the first application allows researchers to draw inferences from a complete RCT while the second application provides researchers with both a recommendation engine to strategically prioritize experiments and a tool for advancing precision medicine. 
%
We further discuss how \SNN~can extract RWE from RWD in four settings: 
(i) synthetic control arms in uncontrolled, single-arm trials, 
(ii-iii) real-world safety and efficacy, 
and (iv) label expansion. 
We leave a formal empirical study of these topics as future work. 
%

%\noindent Please note: Abbreviations should be introduced at the first mention in the main text – no abbreviations lists. Suggested structure of main text (not enforced) is provided below.

%\section*{Introduction}
%
%The Introduction section, of referenced text\cite{Figueredo:2009dg} expands on the background of the work (some overlap with the Abstract is acceptable). The introduction should not include subheadings.
%
%\section*{Results}
%
%Up to three levels of \textbf{subheading} are permitted. Subheadings should not be numbered.
%
%\subsection*{Subsection}
%
%Example text under a subsection. Bulleted lists may be used where appropriate, e.g.
%
%\begin{itemize}
%\item First item
%\item Second item
%\end{itemize}
%
%\subsubsection*{Third-level section}
% 
%Topical subheadings are allowed.
%
%\section*{Discussion}
%
%The Discussion should be succinct and must not contain subheadings.
%
%\section*{Methods}
%
%Topical subheadings are allowed. Authors must ensure that their Methods section includes adequate experimental and characterization data necessary for others in the field to reproduce their work.

\bibliographystyle{alpha}
\bibliography{bib}

\section*{Acknowledgements}
We thank Debra Kientop, Damon Wischik, and Gina Zaghi for their feedback, suggestions, and fruitful discussions. 
%Grant or contribution numbers may be acknowledged.

%\section*{Author contributions statement}
%%
%D.S. conducted the analysis and wrote the manuscript with input from all authors: A.A., V.M., B.S., D.S., H.S., C.W.
%%All authors reviewed the manuscript. 

\section*{Data availability}
Data is under the ownership of TauRx Therapeutics and may be available upon request to B.S. 

\section*{Code availability} 
Code for the \SNN~estimator can be found at 
\href{https://github.com/deshen24/syntheticNN}{https://github.com/deshen24/syntheticNN}. 

%\section*{Additional information}
%%
%The authors declare no competing interests. 

%To include, in this order: \textbf{Accession codes} (where applicable); \textbf{Competing interests} (mandatory statement). 
%
%The corresponding author is responsible for submitting a \href{http://www.nature.com/srep/policies/index.html#competing}{competing interests statement} on behalf of all authors of the paper. This statement must be included in the submitted article file.

\end{document}